\documentclass[a4paper]{article}

\usepackage{itgspeech2025}    
\usepackage{times}            
\usepackage[english]{babel}   
\usepackage[ansinew]{inputenc}
\usepackage[T1]{fontenc}      
\usepackage[sort&compress,numbers]{natbib}	
\usepackage{amsmath,amssymb}
\usepackage{graphicx}
\usepackage[colorlinks=false,pdfborder={0 0 0}]{hyperref}
\usepackage{units}
\usepackage{booktabs}
\usepackage{microtype}
\usepackage[nolist, nohyperlinks]{acronym}
\PassOptionsToPackage{hyphens}{url}
\usepackage{hyperref}  
\begingroup
\makeatletter
\g@addto@macro{\UrlBreaks}{\do\-}
\makeatother
\endgroup
\title{ReverbFX: A Dataset of Room Impulse Responses Derived from Reverb Effect Plugins for Singing Voice Dereverberation}

\author{Julius Richter$^*$, Till Svajda$^*$, Timo Gerkmann}

\address{Signal Processing Group, University of Hamburg, Germany\\
  Email: \texttt{\{julius.richter,timo.gerkmann\}@uni-hamburg.de}}

\begin{acronym}
\acro{DNN}[DNN]{deep neural network}
\acro{STFT}[STFT]{short-time Fourier transform}
\acro{SDE}[SDE]{stochastic differential equation}
\acro{PESQ}[PESQ]{perceptual evaluation of speech quality}
\acro{ESTOI}[ESTOI]{extended short-time objective intelligibility}
\acro{SNR}{signal-to-noise ratio}
\acro{GPU}[GPU]{graphics processing unit}
\acro{RIR}[RIR]{room impulse response}
\acro{FDN}[FDN]{feedback delay network}
\acro{MOS}[MOS]{mean opinion score}
\acrodefplural{MOS}{mean opinion scores}
\end{acronym}

\begin{document}

\maketitle

\begin{abstract}
We present \emph{ReverbFX}, a new room impulse response (RIR) dataset designed for singing voice dereverberation research. 
Unlike existing datasets based on real recorded RIRs, \emph{ReverbFX} features a diverse collection of RIRs captured from various reverb audio effect plugins commonly used in music production. 
We conduct comprehensive experiments using the proposed dataset to benchmark the challenge of dereverberation of singing voice recordings affected by artificial reverbs. 
We train two state-of-the-art generative models using \emph{ReverbFX} and demonstrate that models trained with plugin-derived RIRs outperform those trained on realistic RIRs in artificial reverb scenarios. 
\end{abstract}

\renewcommand{\thefootnote}{\fnsymbol{footnote}} 
\footnotetext[1]{Authors contributed equally to this work.} 
\renewcommand{\thefootnote}{\arabic{footnote}} 

\section{Introduction}

\Acp{RIR} play a crucial role in audio signal processing, as they characterize how sound propagates and decays in a given environment~\cite{pierce2019acoustics}. 
To obtain \acp{RIR} that characterize natural reverberation, they can either be measured in actual rooms or simulated using established acoustic models~\cite{scheibler2018pyroomacoustics}.
These \acp{RIR} reflect the physical properties and geometries of actual spaces, yielding reverberation with consistent time decay, energy distribution, and frequency response characteristics. 
Accordingly, most \ac{RIR} datasets and dereverberation benchmarks are based on this natural room acoustics paradigm, aiming to improve technologies such as speech recognition, hearing aids, and audio enhancement in real-world environments~\cite {naylor2010speech}.

However, there is a discrepancy between natural reverberation and artificial reverb effects used in modern music production~\cite{moorer1979reverberation, valimaki2012fifty}.
Unlike real-world acoustics, music production often relies on digital reverb effects that are directly applied to individual tracks, such as vocals, using software plugins.
These plugins can employ convolution-based methods using recorded or synthesized \acp{RIR}, or algorithmic approaches that simulate reverberation through mathematical models. 
Artificial reverbs are unconstrained by the laws of physics and can exhibit a wide range of exotic behaviors: extremely long or modulated decays, non-monotonic frequency responses, non-linear characteristics, and even deliberately unnatural acoustics. 
These reverberation effects are part of the creative palette, shaping the aesthetics of contemporary music. 
As a consequence, dereverberation approaches trained and evaluated exclusively on natural \ac{RIR} datasets may not generalize well to these artificial reverberation scenarios.

To address this gap, we introduce \emph{ReverbFX}, a novel dataset of \acp{RIR} entirely generated from a diverse set of professional reverb audio effect plugins. 
Unlike existing \ac{RIR} datasets, \emph{ReverbFX} captures the diversity and complexity of artificial reverbs frequently used in music production. 
This unique dataset enables, for the first time, the study and benchmarking of dereverberation models specifically for these artificial environments.

We use \emph{ReverbFX} to create a new benchmark, \emph{Singing\-ReverbFX}, for evaluating singing voice dereverberation under artificial reverberation conditions. 
Using this data, we train two state-of-the-art generative models \cite{richter2023speech, jukic24_interspeech} and demonstrate that, when trained on \emph{SingingReverbFX}, these models can effectively dereverberate singing voices recorded with artificial reverbs, significantly outperforming counterparts trained solely on natural RIRs (i.e., in an off-domain setting). 
Our experiments establish \emph{ReverbFX} as an essential resource for advancing dereverberation research beyond natural acoustics and towards applications in music production, where artificial reverbs predominate.

In summary, our contributions are three-fold:
\begin{enumerate}
    \item We motivate and construct \emph{ReverbFX}, the first dataset focused on artificial reverberation from audio effect plugins.
    \item We systematically compare two generative dereverberation models trained on artificial versus natural reverberation.
    \item We establish a new benchmark for singing voice dereverberation in artificial reverb scenarios, paving the way for more robust and musically relevant audio enhancement technologies.
\end{enumerate}

\section{Background}

Reverberation, or reverb for short, is a phenomenon that occurs in enclosed spaces when sound signals reflect off walls and other surfaces~\cite{naylor2010speech}. 
In music production, reverb is one of the most commonly used effects, applied to almost every track in a mix. 
This is especially true for vocals, where reverb is used to create tension, add texture, or simply achieve a more natural and immersive sound. 
Reverb can make vocals sound wider and significantly influence the emotional response of listeners; for example, music with long, dense reverb often evokes an epic or dreamlike atmosphere.

Music datasets often provide audio in separate stems, such as vocals, drums, bass, and accompaniment. These stems can be used for training source separation models~\cite{rouard2023hybrid}, stem mixing and mastering~\cite{woodhall2010audio}, and other music information retrieval tasks~\cite{lerch2022introduction}.
However, vocal stems are often provided with added reverb, and truly anechoic (dry) voice recordings are not always available. 
Therefore, singing voice dereverberation is necessary to retrieve the original anechoic voice. In turn, the resulting dry singing voice can be convolved with a variety of \acp{RIR} to generate additional data for training purposes--a common data augmentation strategy.
Moreover, in music production, it is usually preferable to apply reverb to non-reverberant vocals, as this allows precise parameter control and greater creative flexibility.


Removing reverb from a signal, especially from vocals or speech, is a challenging problem in the field of signal processing. Singing voice dereverberation is conceptually similar to speech dereverberation, and similar methods can be used to address both tasks. However, despite their similarities, speech and singing voice have distinct characteristics, particularly in terms of pitch, power, and phoneme duration~\cite{sundberg2018singing}.

Early approaches to vocal and music dereverberation primarily rely on signal processing techniques such as linear prediction \cite{musicDereverbWSpectralLinearPrediction, musicDereverbWNonparametricBayesian, musicDereverbWmultiChannelLinPre, nakatani2008speech}, or source models designed to capture the harmonic structure of musical signals \cite{MusicDereverbWHarmonics, musicDereverbWIDivergence}. 
These methods typically assume a structured representation of the source and exploit music's inherent periodicity and tonal characteristics to mitigate reverberation effects.

Recent research in machine learning has introduced a variety of new approaches, including end-to-end deep neural networks~\cite{koo2021reverb}. 
These supervised methods tend to learn the characteristics of specific reverbs present in the training data. This leads to reduced performance when encountering unseen reverb types, a limitation given the wide range of possible reverberation scenarios in real-world applications. 
To address this challenge, more recent work has increasingly focused on generative approaches, such as diffusion models, which can be supervised or unsupervised and aim to achieve robust generalization across diverse reverb conditions~\cite{lemercier2025diffusion}.

\paragraph{Feedback Delay Networks}

One of the foundational concepts behind algorithmic reverb plugins was introduced by Stautner and Puckette in 1982~\cite{fdn}. 
Their approach feeds multiple copies of the input signal into a network of delay lines with varying lengths. 
The outputs from these delays are then weighted and recursively routed back into the system through a feedback delay matrix, laying the groundwork for what would later be formalized as \acp{FDN}~\cite{zolzer2002dafx}. 
This architecture allows the signal to circulate within the network, mimicking the complex reflections in natural acoustic environments. 
The feedback matrix is typically unitary to ensure filter stability, resulting in a lossless \ac{FDN}. 
The reverb tail's envelope is generally shaped using a combination of gain-based decay and low-pass filtering on the delay lines, simulating the gradual energy loss, especially of high frequencies, that occurs with multiple reflections in real spaces.  
   
Beyond the feedback process, some algorithms also include a feedforward path that routes the input signal through the delay lines once before reaching the output. 
This component helps approximate early reflections and provides additional dry/wet balance control.

\begin{figure}[t]
	\centerline{\includegraphics[width=.95\columnwidth]{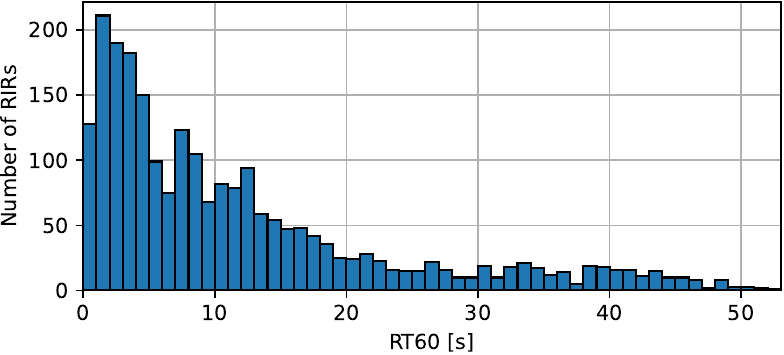}}
    \vspace{-0.5em}
	\caption{Histogram of RT60 values for \emph{ReverbFX} dataset, showing the distribution of reverberation times among the collected \acp{RIR}.}
	\label{fig:histogram}%
\end{figure}

\section{ReverbFX Dataset}

We release the \emph{ReverbFX} dataset\footnote{\url{https://sp-uhh.github.io/reverbfx}}, comprising 1,846 \acp{RIR}, with RT60 times ranging from 0.31 to 52.08 seconds. 
A histogram of RT60 times is shown in Figure~\ref{fig:histogram}.
All \acp{RIR} are generated using four publicly available audio plugins (see Section~\ref{sec:plugins})  and measured using a discrete-time unit impulse as the input signal.
The reverb effects are applied via \emph{DawDreamer}~\cite{dawdreamer}, a Python API providing DAW functionality, including support for VST plugins.

For each plugin, we utilize both factory and custom presets. 
To further increase variation in the resulting audio effects, we generate 14 randomized versions of each preset by modifying two randomly selected parameters from a predefined subset (e.g., delay time and low cut). 
This procedure results in a total of 15 impulse responses per preset. 
For all configurations, the dry/wet parameter is set to 1 to ensure a fully wet signal.

A rigorous validation procedure is employed to filter out anomalous or corrupted data to guarantee the quality and reliability of the \acp{RIR}. Each \ac{RIR} underwent the following checks:
\begin{itemize}
    \item \textbf{Numerical Integrity:} Verification for invalid numerical values such as NaNs or infinite entries.
    \item \textbf{DC Offset:} Ensuring the mean value is close to zero (threshold:  $10^{-4}$)
    \item \textbf{Energy Threshold:} Discarding RIRs with an energy below $10^{-16}$
    \item \textbf{Length Validation:} Excluding empty or zero-length RIRs.
    \item \textbf{RT60 Consistency:} Retaining only those RIRs with a computed RT60 in the range of 0.1 to 100 seconds.
    \item \textbf{Hilbert Envelope:} Evaluating the amplitude envelope via the Hilbert transform to exclude RIRs with excessively short decay times.
\end{itemize}

\begin{table*}[t]
\centering
\begin{tabular}{lrcrllcc}
\toprule
\textbf{Dataset} & \textbf{Duration} & \textbf{\#$\,$Singers} & \textbf{\#$\,$Files}  & \textbf{Song type} & \textbf{Language} & $f_s$ & \textbf{Bit depth} \\
& \scriptsize{(Hours)} & & & & & \scriptsize{(kHz)} & \scriptsize{(bit)} \\
\midrule
OpenSinger~\cite{opensinger} & $51.9$ & $25$ m, $41$ f & $39116$ & Traditional and Pop Songs & Chinese & $24$ & $16$ \\
M4Singer~\cite{M4Singer} & $29.8$ & $10$ m, $10$ f & $20696$ & Pop Songs & Mandarin & $44.1$ & $24$ \\
CSD~\cite{CSD}  & $4.7$ & $1$ f & $1810$ & Children Songs & Korean, English & $44.1$ & $16$ \\
PJS~\cite{PJS} & $0.5$ & $1$ m & $195$ & Children and Anime Songs & Japanese & $48$ & $24$ \\
Opencpop~\cite{opencpop}  & $5.2$ & $1$ f & $2517$ & Chinese Pop Songs & Chinese & $44.1$ & $24$ \\
NUS-48E~\cite{NUS48} & $1.9$ & $6$ m, $6$ f & $703$ & Traditional and Pop Songs & English & $44.1$ & $16$ \\
\midrule
NHSS~\cite{sharma2021nhss} \scriptsize{(Test)} & $4.8$ & $5$ m, $5$ f & $490$ & Pop Songs & English & $48$ & $16$ \\
\bottomrule
\end{tabular}%
\caption{Overview of vocal datasets used for training and the test. 
The number of speakers is indicated by their gender, male (m) or female (f), and $f_s$ denotes the sample rate. The number of files reflects only those included in the dataset and does not count files discarded during the generation process.}
\label{tab:datasets}
\end{table*}

\subsection{Reverb Audio Effect Plugins}\label{sec:plugins}

\paragraph{Protoverb (v1.0.1)} 
The Protoverb plugin is proprietary software distributed by u-he Heckmann Audio GmbH\footnote{\url{https://u-he.com}} and licensed for free, non-commercial use. 
Contrary to traditional algorithmic reverbs that aim to minimize resonances, Protoverb builds up as many room resonances as possible, modeling the body of air in a room~\cite{protoverb}. It is based on a delay network, whose structure is controlled by a text-based randomization feature, allowing users to generate unique reverb settings. 
For our study, we created $200$ presets with different random strings in the reverb section to capture a diverse set of \acp{RIR}.

\paragraph{SkyNet (v0.4)}
SkyNet Reverb by Shift Line\footnote{\url{https://shift-line.com}} is proprietary software offered under a Pay-What-You-Want model. It is based on the two reverb algorithms SkyNet-1 and SkyNet-2 from the Astronaut space reverb pedal, and is designed to create expansive soundscapes with long decay times and diffuse sound~\cite{skynet}. The plugin offers a parameter to control the mix ratio between the two algorithms.
 
\paragraph{TAL-Reverb-4 (v4.0.4)}
Tal-Reverb-4 is a proprietary software distributed by TAL Software GmbH\footnote{\url{https://tal-software.com}}.
It is based on a modulated delay network and is designed to evoke the character of 1980s digital reverbs~\cite{tal_reverb_4}. The algorithm produces a high-quality, plate-style reverb with a diffuse and modulated character, making it particularly suitable for vocals and synthetic sounds.

\paragraph{Valhalla Supermassive (v3.0.0)}
The Valhalla Supermassive plugin is proprietary software distributed by Valhalla DSP\footnote{\url{https://valhalladsp.com}} and is offered as freeware. 
Designed for extreme and lush reverb and delay effects, it implements a modulated \ac{FDN} architecture optimized for massive spatialization, long decay times, and evolving textures~\cite{valhalla_supermassive}. The plugin offers a range of algorithms, each with unique temporal and spectral characteristics, making it particularly suited for ambient sound design and experimental processing. 

\subsection{License}

We obtained explicit permission from the plugin developers to generate and publish \acp{RIR} derived from the plugins for non-commercial research purposes. 
The released dataset does not contain any part of the original software. 
\emph{ReverbFX} is released under the \href{https://creativecommons.org/licenses/by-nc/4.0/}{Creative Commons Attribution-Non\-Commercial 4.0 International License (CC BY-NC 4.0)}. It may be used, shared, and adapted for non-commercial research and educational purposes only. 
Commercial use is strictly prohibited. 
Proper attribution must be given to the dataset authors and the original plugin developers.

\section{SingingReverbFX Benchmark}

We introduce \emph{SingingReverbFX}, a singing voice dereverberation benchmark focused on restoring reverberant vocals processed with audio reverb effects commonly used in music production.
To this end, we convolve singing voice data with our new \emph{ReverbFX} dataset of \acp{RIR} captured from reverb effect plugins.

\subsection{Data} 

Inspired by Lemercier et al.~\cite{BUDDy}, we utilize a variety of singing voice datasets, including \emph{OpenSinger}~\cite{opensinger}, \emph{M4Sing\-er}~\cite{M4Singer}, \emph{CSD}~\cite{CSD}, \emph{PJS}~\cite{PJS}, \emph{OpenCpop}~\cite{opencpop}, \emph{NUS-48E}~\cite{NUS48}, and NHSS~\cite{sharma2021nhss}.
These datasets feature studio-quality recordings with balanced gender and phoneme distributions and include vocals in multiple languages. 
The combined vocal data amounts to a total of 94 hours. 
An overview of the datasets and their key characteristics is provided in Table~\ref{tab:datasets}.

To prepare data for training, validation, and the test, we partition the \emph{ReverbFX} dataset using stratified sampling based on RT60 bins, ensuring that each subset is representative of the overall distribution of reverberation times. 
The data is split into training, validation, and test sets with a ratio of 1446:200:200. 
The vocal training set includes all recordings except those set aside for validation and testing. 
For validation, we select 980 gender-balanced files from the OpenSinger dataset. 
The entire NHSS dataset is used exclusively for testing.

For every split, each vocal recording is paired with a randomly selected \ac{RIR} from the corresponding subset. 
To ensure compatibility for convolution, vocal files are first resampled to 48~kHz. 
Each \ac{RIR} is truncated at its maximum index to remove trailing silence or noise, then normalized, converted to mono by taking the left channel, and convolved with the vocal signal.

After convolution, we apply dry/wet mixing, a standard parameter in digital reverb plugins, which defines the balance between the original (dry) and reverberated (wet) vocals. 
At each iteration, the mixing ratio $\alpha$ is randomly sampled from a uniform distribution $\alpha \in [0.1,1]$, resulting in
\begin{equation}
    y(t) = (1-\alpha) x(t) + \alpha (x(t) \ast h(t)),
\end{equation}
where 
$y(t)$ is the final mixture, $x(t)$ is the dry vocal, and $h(t)$ is the \ac{RIR}.

Our approach, convolving vocals with \acp{RIR}, leverages the principles of linearity and time invariance (LTI), meaning that only linear, time-invariant reverb characteristics can be captured. 
Some reverb plugins, however, include time-varying or nonlinear elements (e.g., Chorus or Pitch Modulation) that cannot be replicated through \ac{RIR} convolution. 
Nevertheless, using \acp{RIR} offers a straightforward way to construct the dataset while still modeling essential reverb parameters, such as delay time, intensity, and absorption characteristics.

\begin{table*}[t]
\centering
\begin{tabular}{lcccccc}
\toprule
 & \textbf{POLQA} & \textbf{PESQ} & \textbf{ESTOI} & \textbf{MAE} $\downarrow$ & \textbf{OVRL} & \textbf{REVERB} \\
\midrule
Reverberant & $1.89 \pm 1.01$ & $1.61 \pm 0.76$ & $0.41 \pm 0.29$ & $0.14 \pm 0.12$ & $1.81 \pm 0.69$ & $3.11 \pm 0.58$ \\
\midrule
\emph{SGMSE+}~\cite{richter2023speech} (mism.) & $2.53 \pm 1.31$ & $2.23 \pm 1.06$ & $0.56 \pm 0.27$ & $0.14 \pm 0.10$ & $2.72 \pm 0.47$ & $4.22 \pm 0.39$ \\
\emph{SGMSE+}~\cite{richter2023speech} & $\mathbf{3.12 \pm 1.33}$ & $\mathbf{2.63 \pm 1.10}$ & $\mathbf{0.64 \pm 0.24}$ & $\mathbf{0.08 \pm 0.06}$ & $\mathbf{2.99 \pm 0.32}$ & $\mathbf{4.30 \pm 0.38}$ \\
\emph{Schr\"{o}dinger Bridge}~\cite{jukic24_interspeech} & $2.90 \pm 1.25$ & $2.37 \pm 1.05$ & $0.61 \pm 0.24$ & $0.13 \pm 0.09$ & $2.85 \pm 0.42$ & $4.12 \pm 0.46$ \\
\bottomrule
\end{tabular}
\caption{Singing voice dereverberation results on \emph{SingingReverbFX} using standard speech enhancement metrics. Values indicate mean and standard deviation. Higher values indicate better performance for all metrics except MAE.}
\label{tab:results}
\end{table*}

\subsection{Baselines}

All models are trained under identical conditions using two NVIDIA RTX A6000 \acp{GPU}, with a batch size of 6. Training is stopped after 350$\,$k steps. 
The following paragraphs provide details about the specific models used in our experiments.

\paragraph{Score-based Generative Models}
Richter et al.~\cite{richter2023speech} introduced \emph{SGMSE+}, a score-based generative model for speech enhancement and dereverberation. In a supervised setting, \emph{SGMSE+} learns clean speech posteriors conditioned on noisy or reverberant inputs. The model operates in the complex \ac{STFT} domain, utilizing a diffusion process and a neural network to estimate the score function, which guides the reverse generative process. 

In a follow-up study~\cite{richter2024ears}, the authors adapted the model for 48\,kHz audio and conducted an extensive hyperparameter search to optimize the noise schedule and amplitude compression of the input representation. 
We adopt these optimal hyperparameters to train \emph{SGMSE+} on \emph{SingingReverbFX} using the official implementation.\footnote{\url{https://github.com/sp-uhh/sgmse}}
As an ablation, we also train \emph{SGMSE+} on natural \acp{RIR} using the datasets employed in~\cite{richter2024ears} to illustrate the domain shift between natural and artificial reverberation.

\paragraph{Schr\"{o}dinger Bridge}

Juki\'{c} et al.~\cite{jukic24_interspeech} extended \emph{SGMSE+} by employing a tractable Schr\"{o}dinger bridge as the diffusion process for speech enhancement and dereverberation. 
The Schr\"{o}dinger bridge transforms the input speech distribution into the clean speech distribution by formulating and solving an optimal transport problem. 
Richter et al.~\cite{richter2024diffusion, richter2025investigating} later adapted this model for 48\,kHz audio and integrated it into their code repository, which we use for training the Schr\"{o}dinger bridge on \emph{SingingReverbFX}. 
We utilize the \ac{SDE} sampler, as it has been shown to provide superior performance for dereverberation tasks~\cite{jukic24_interspeech}.

\subsection{Metrics}

Currently, there are no metrics specifically designed for singing voice dereverberation. Therefore, we adopt a set of established metrics from speech enhancement to evaluate both perceptual quality, intelligibility, and signal fidelity:
\begin{itemize}
    \item \textbf{POLQA}~\cite{beerends2013perceptual}: An intrusive, model-based standard (ITU-T P.863) for perceptual audio quality, considered a successor to PESQ and widely used in speech evaluation.
    \item \textbf{PESQ}~\cite{PESQ}: An intrusive metric that estimates perceived speech quality on a scale from 1 to 4.5, originally developed for telephony.
    \item \textbf{ESTOI}~\cite{ESTOI}: Measures intelligibility by comparing short-time spectral features between the reference and output, giving a score from 0 (unintelligible) to 1 (perfect).
    \item \textbf{MAE}: Mean absolute error in the \ac{STFT} domain, providing a direct, objective comparison between estimated and reference signals.
    \item \textbf{OVRL}~\cite{naderi2024multi}: A non-intrusive neural network-based estimator that predicts \ac{MOS} for overall quality.
    \item \textbf{REVERB}~\cite{naderi2024multi}: A non-intrusive neural network-based estimator that predicts \ac{MOS} for perceived reverberation.
\end{itemize}
Together, these metrics provide a comprehensive evaluation framework to assess both the perceptual and objective quality of dereverberated singing voice signals.

\section{Results}

In Table~\ref{tab:results}, we present the singing voice dereverberation results obtained using \emph{SingingReverbFX}.  
It can be seen that all methods yield improvements over the reverberant test set across all metrics.  
Furthermore, training \emph{SGMSE+} on natural reverb (with mismatched \acp{RIR}) leads to a performance decrease compared to training \emph{SGMSE+} with in-domain data from the \emph{SingingReverbFX} training set.
The \emph{Schr\"{o}dinger bridge} also demonstrates strong performance but is slightly outperformed by \emph{SGMSE+}. This may be attributed to the extensive hyperparameter tuning applied to the 48
kHz model in \cite{richter2024ears}.

\section{Conclusions}

In this work, we introduced \emph{ReverbFX}, the first dataset dedicated to artificial reverberation effects commonly used in music production. 
Through comprehensive benchmarking with state-of-the-art generative models, we demonstrated that models trained on plugin-derived \acp{RIR} significantly outperform those trained on traditional, natural \acp{RIR} when faced with dereverberation of singing voices under artificial reverb conditions. 
Our results highlight the importance of domain-specific data for dereverberation tasks and establish \emph{ReverbFX} and the \emph{Singing-ReverbFX} benchmark as vital resources for developing and evaluating audio enhancement models tailored to modern music production environments.

\small
\bibliographystyle{ieeetr}
\bibliography{refs}


\end{document}